\documentclass{aa}  
\usepackage{amssymb,epsfig,graphicx}
\usepackage{txfonts}
\begin{document}

\title{Probing Cosmic Chemical Evolution with Gamma-Ray Bursts:
  GRB\,060206 at z = 4.048 \thanks{Based on observations made at the
    Spanish Observatorio del Roque de los Muchachos on La Palma with
    the Nordic Optical Telescope, operated jointly by Denmark,
    Finland, Iceland, Norway, and Sweden, and with the William
    Herschel Telescope.}}

\author{
J.~P.~U.~Fynbo \inst{1} \and
R.~L.~C.~Starling \inst{2} \and
C.~Ledoux \inst{3} \and
K.~Wiersema \inst{2} \and
C.~C.~Th{\"o}ne \inst{1}\and
J.~Sollerman \inst{1}\and
P.~Jakobsson \inst{1} \and
J.~Hjorth \inst{1} \and
D.~Watson \inst{1} \and
P.~M.~Vreeswijk \inst{4,3} \and 
P.~M\o ller \inst{5} \and
E.~Rol \inst{6} \and
J.~Gorosabel \inst{7} \and
J.~N{\"a}r{\"a}nen \inst{8} \and
R.~A.~M.~J.~Wijers \inst{2} \and
G.~Bj{\"o}rnsson \inst{9} \and
J.~M.~Castro~Cer\'on \inst{1} \and
P.~Curran \inst{2} \and
D.~H.~Hartmann \inst{10} \and
S.~T.~Holland \inst{11} \and
B.~L.~Jensen \inst{1} \and
A.~J.~Levan  \inst{12} \and
M.~Limousin \inst{1} \and
C.~Kouveliotou \inst{13} \and
G.~Nelemans \inst{14} \and
K.~Pedersen \inst{1} \and
R.~S.~Priddey \inst{12} \and
N.~R.~Tanvir \inst{12} 
}

\institute{
	  Dark Cosmology Centre, Niels Bohr Institute, University of
	  Copenhagen, \mbox{Juliane Maries Vej 30, 2100 Copenhagen, Denmark}
	  \and 
	  Astronomical Institute `Anton Pannekoek', University of Amsterdam,
	  Kruislaan 403, 1098 SJ Amsterdam, the Netherlands
	  \and
	  European Southern Observatory, Alonso de C\'{o}rdova 3107, 
	  Casilla 19001, Vitacura, Santiago, Chile
	  \and
	  Departamento de Astronom\'ia, Universidad de Chile, Casilla 36-D, 
	  Santiago, Chile
	  \and
	  European Southern Observatory, Karl-Schwarzschild-strasse 2,
	  D-85748 Garching bei M{\"u}nchen, Germany
	  \and
	  Department of Physics and Astronomy, University of Leicester, 
	  University Road, Leicester LE1 7RH, UK
	  \and
	  Instituto de Astrof\'{\i}sica de Andaluc\'{\i}a (CSIC), 
	  Apartado de Correos 3004, E-18080 Granada, Spain
	  \and
	  Observatory, University of Helsinki, PO Box 14,
	  FIN-00014 Helsinki, Finland
	  \and 
	  Science Institute, University of Iceland, Dunhaga 3, i
	  107 Reykjav\'{\i}k, Iceland
	  \and
	  Department of Physics and Astronomy, Clemson University, Clemson, 
	  South Carolina 29634-0978, USA
	  \and
	  NASA Goddard Space Flight Center, Greenbelt, MD 20771, USA
	  \and
	  Centre for Astrophysics Research, University of Hertfordshire,
	  College Lane, Hatfield, Hertfordshire AL10 9AB, UK
	  \and
	  NASA Marshall Space Flight Center, NSSTC, XD-12, 
	  320 Sparkman Drive, Huntsville, AL 35805, USA
	  \and
	  Department of Astrophysics, Radboud University, 
	  PO Box 9010, 6500 GL Nijmegen, the Netherlands
}


\date{Received 2006 / Accepted 2006}

\abstract {} {We present early optical spectroscopy of the afterglow
  of the gamma-ray burst GRB\,060206 with the aim of determining the
  metallicity of the GRB absorber and the physical conditions in the
  circumburst medium. We also discuss how GRBs may be important
  complementary probes of cosmic chemical evolution.}  {Absorption
  line study of the GRB afterglow spectrum.}  {We determine the
  redshift of the GRB to be $z=4.04795\pm0.00020$. Based on the
  measurement of the neutral hydrogen column density from the damped
  Lyman-$\alpha$ line and the metal content from weak, unsaturated
  \ion{S}{ii} lines we derive a metallicity of
  [S/H]$=-0.84\pm0.10$. This is one of the highest metallicities
  measured from absorption lines at $z\sim4$. From the very high
  column densities for the forbidden \ion{Si}{ii*}, \ion{O}{i*},
  and \ion{O}{i**} lines we infer very high densities and low
  temperatures in the system. There is evidence for the presence
  of H$_2$ molecules with log~$N$(H$_2$)$\sim$17.0, translating
  into a molecular fraction of $\log{f}\approx -3.5$ with
  $f=2N($H$_2$)/(2$N$(H$_2$)+ $N(\ion{H}{i})$). Even if GRBs are
  only formed by single massive stars with metallicities below
  $\sim0.3~Z_{\odot}$, they could still be fairly unbiased tracers
  of the bulk of the star formation at $z>2$.  Hence,
  metallicities as derived for GRB\,060206 here for a complete
  sample of GRB afterglows will directly show the distribution of
  metallicities for representative star-forming galaxies at these
  redshifts.}  {}


\titlerunning{Probing Cosmic Chemical Evolution with Gamma-Ray Bursts}

\authorrunning{J. P. U. Fynbo et al.}

\maketitle

\section{Introduction}

Long gamma-ray bursts (GRBs) are now established to be caused by the deaths of
massive stars (e.g., \cite{jensNature}, \cite{stanek03}) and due to their
brightness they can be observed throughout most of the observable Universe
(e.g., \cite{Kawai}). Given these facts it has long been realized that GRBs
could be ideal probes of star-formation activity throughout the history of the
Universe (e.g., \cite{wijers98}).  However, previous GRB missions detected too
few rapidly well-localised GRBs to build statistically interesting samples and
hence really capitalize on this potential. The {\it Swift} satellite (Gehrels
et al.~\cite{gehrels}) has increased the detection rate of rapidly
well-localised GRBs by roughly an order of magnitude compared to previous
missions. Moreover, its significantly deeper detection limit (e.g., Band~
\cite{band}) means that {\it Swift} detects more distant bursts than previous
missions (\cite{palli06}).  

In this {\it Letter} we present optical spectroscopy of GRB\,060206,
focussing on the measurement of the metallicity of the GRB absorption
system, the density in the circumburst environment and the presence of
molecules.  We then discuss how GRBs are important for understanding
the build-up and distribution of metals in galaxies (e.g., Pei \&
Fall~\cite{peifall}), which is currently not fully understood (e.g.,
Ferrara et al.~\cite{ferrara}).

\section{Observations}
\label{obs.sec}

GRB\,060206 was discovered by the Burst Alert Telescope (BAT) aboard the {\it
Swift} satellite on February 6 04:46:53 UT.  The burst exhibited a slow rise
and a faster decline, with a $T_{90}$ of $7\pm2$ s (Palmer et
al.~\cite{Palmer}).  The X-ray Telescope (XRT) slewed prompt\-ly to the
location and began taking data at \mbox{$\Delta t = 58$\,s}, where $\Delta t$
is the time from the onset of the burst. Due to entry into the South Atlantic
Anomaly, XRT could only observe the BAT error circle briefly, and therefore no
fading X-ray source was immediately localised (Morris et al.~\cite{morris}).
Observations with the UV/Optical Telescope (UVOT) began at \mbox{$\Delta t =
57$\,s} but failed to reveal an optical afterglow (OA) candidate in the initial
data products (Morris et al.~\cite{morris}).  \par We observed the GRB\,060206
BAT error circle in the $R$-band with the Andaluc\'{\i}a Faint Object
Spectrograph and Camera (ALFOSC) on the Nordic Optical Telescope (NOT) starting
at \mbox{$\Delta t \approx 15$\,min}. A point-like object ($R\sim17.3$) not
present in the Digitized Sky Survey was detected. The detection was confirmed by
re-analysis of the XRT and UVOT data (Boyd et al.~\cite{boyd}; Morris et
al.~\cite{morrisb}). 

Starting at \mbox{$\Delta t \approx 48$\,min} we obtained a
1800\,s spectrum with a low resolution (LR) grism and a 1.3 arcsec wide slit
covering the spectral range from about 3500 \AA \ to 9000~\AA \ at a resolution
of 14 \AA. The airmass during the spectroscopic observation was very low, 
starting at 1.01. 
We also obtained medium resolution (MR) spectroscopic data with
the Intermediate-dispersion Spectroscopic and Imaging System (ISIS) on the
William Herschel Telescope (WHT) at $\Delta t = 1.61$ hours.
We took spectra of exposure times 2$\times$900 s with
both the the 600B grating (covering 3800--5300 \AA) and and the 1200R grating
(covering 6100--7200 \AA) both with a 1.0 arcsec wide slit. The mean airmass
was 1.036 and the observations were carried out at the parallactic angle.  The
resolutions of blue- and red-arm spectra are 1.68 and 0.82 \AA, respectively.

\section{Results}

\begin{figure}
\centering
\epsfig{file=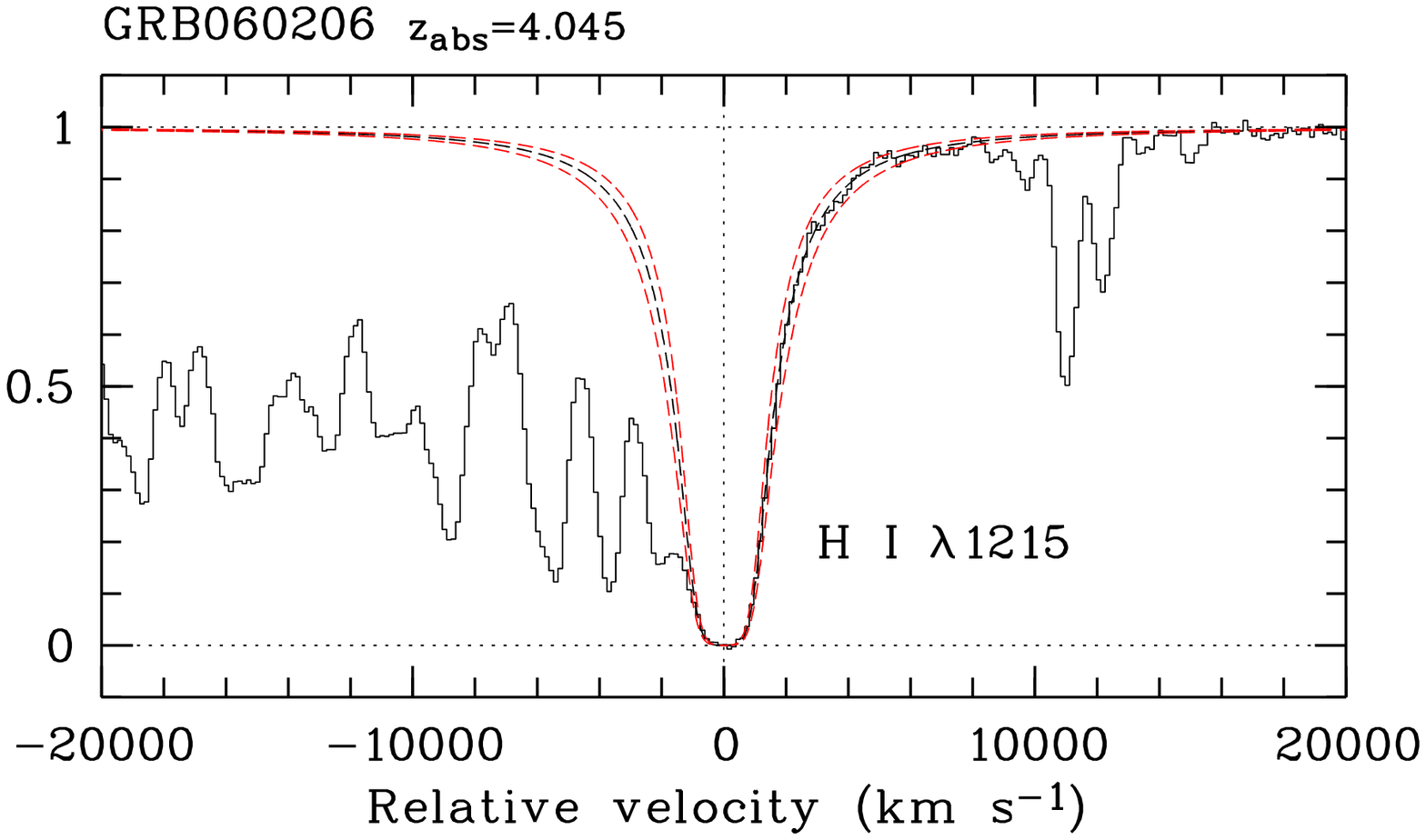,width=8.5cm,clip=}
\epsfig{file=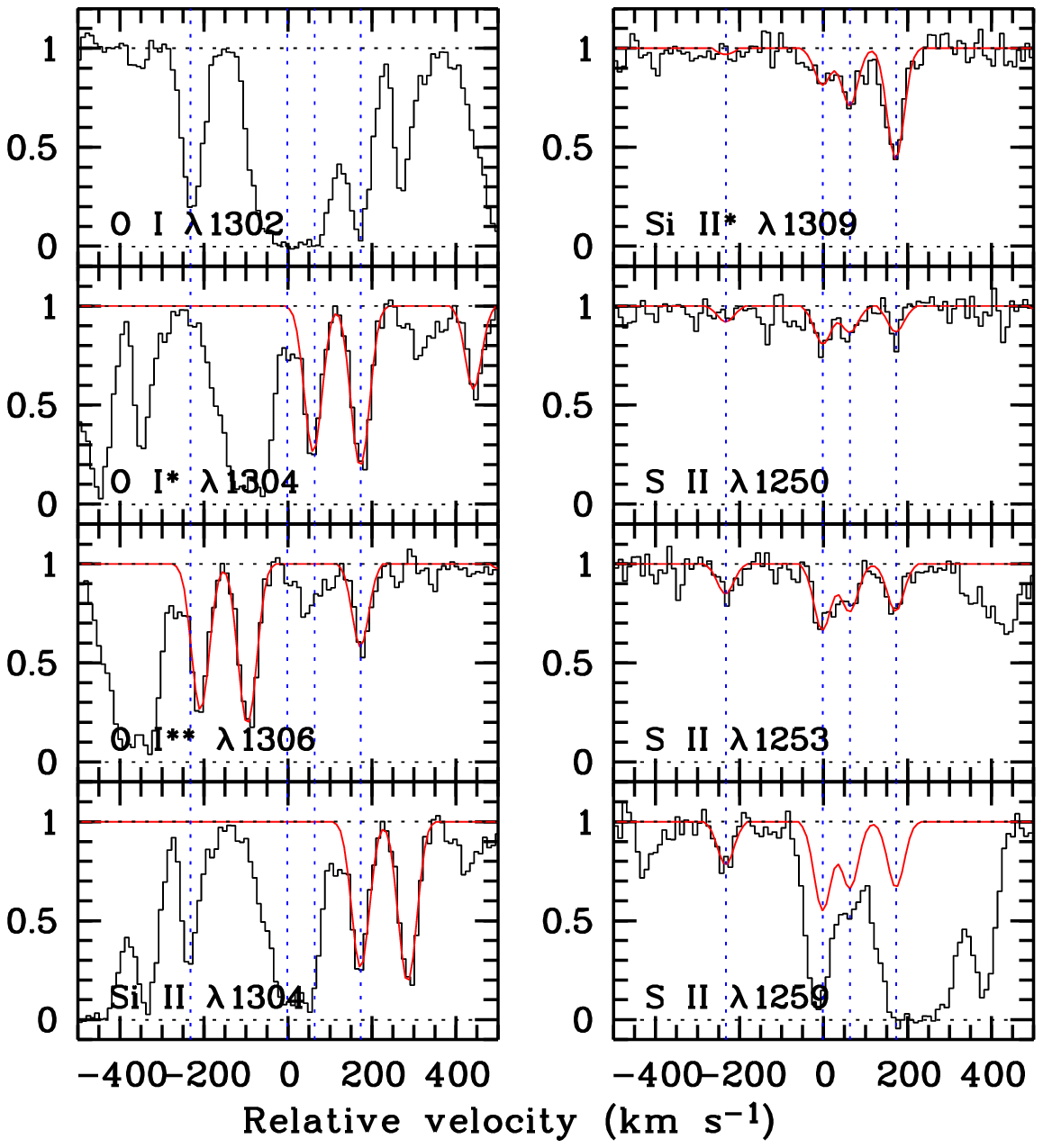,width=8.5cm,clip=}
\epsfig{file=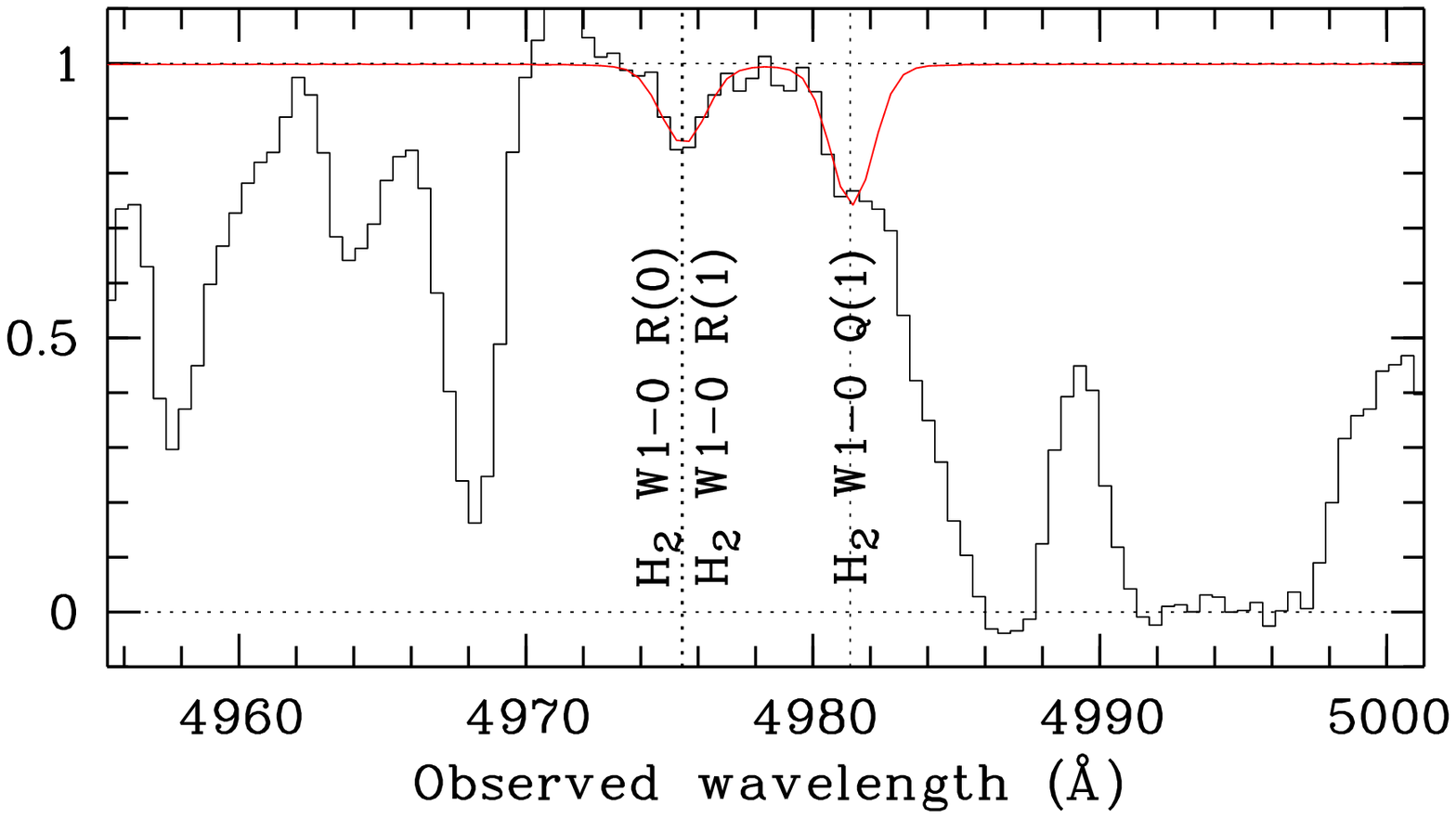,width=8.5cm,clip=}
\caption{{\it Upper panel:} Section of the LR afterglow spectrum
  showing the DLA line at the GRB redshift, $z_{\rm
    abs}=4.048$. Overlaid is the best fitting DLA profile,
  corresponding to $\log N$(\ion{H}{i}) $=$ 20.85 $\pm$ 0.10.  {\it
    Middle panel:} Fits to the \ion{O}{i}, \ion{O}{i*}, \ion{O}{i**},
  \ion{Si}{ii}, \ion{Si}{ii*}, and \ion{S}{ii} lines in the MR
  spectrum. The zero-point for the velocity
  scale is $z_{\rm abs}=4.048$.  {\it Lower panel:} The likely H$_2$
  lines at the GRB redshift in the MR spectrum.
\label{LRMRspectrum}} 
\end{figure}

The spectra show absorption both at the GRB redshift of
$z=4.04795\pm0.00020$ and from two intervening systems at redshifts
$z=1.48$ and $z=2.26$. In the analysis presented here we focus on the
metallicity of the GRB absorption system.  From a fit of the damped
Ly$\alpha$ line at $z=4.048$, we measure a neutral hydrogen column
density of $\log N$(\ion{H}{i}) $=$ 20.85 $\pm$ 0.10
(Fig.~\ref{LRMRspectrum}), well above the definition for Damped
Ly$\alpha$ Absorbers (DLAs, $\log N$(\ion{H}{i}) $\ge$ 20.3). This
value is consistent with fits to the higher order Lyman lines in the
blue MR spectrum. To derive the metal column density we use the
unsaturated \ion{S}{ii} $\lambda$1250, 1253, 1259 lines shown in
Fig.~\ref{LRMRspectrum}.  The profile of the GRB absorption systems
consists of at least four components spread over $\sim$500 km s$^{-1}$
in velocity space. On the right hand side of the middle panel of
Fig.~\ref{LRMRspectrum}, a four-component fit of \ion{S}{II} and
\ion{Si}{II*} is shown. On the left hand side, we have only fitted the
redmost component at 175 km s$^{-1}$ where we can derive a column
density for \ion{Si}{II}, \ion{O}{I*} and \ion{O}{I**}. We fit the
lines using the FitLyman package in MIDAS; the measured column
densities for all components are given in Table~\ref{metallines}.
Note that for components 2 and 3 we fixed the turbulent broadening
parameter value to 15 km s$^{-1}$. Summing over the four components of
the system, we get $\log N$(\ion{S}{II}) $= 15.21\pm0.03$ which leads
to [S/H] $=-0.84\pm0.10$.

We also find $\log N$(\ion{Si}{II*}) $= 14.42\pm0.02$ and therefore
[\ion{Si}{II*}/\ion{Si}{II}]$ = -1.15$, assuming [S/Si] = 0. In the
fourth component, we have [\ion{Si}{II*}/\ion{Si}{II}] $=-0.34$, which
is high compared to the ratios found in other GRB host galaxies (twice
as high as the value found for GRB\,050505 by Berger et
al.~\cite{berger} and 20 times higher than the value found for
GRB\,030323 by Vreeswijk et al.~\cite{vreeswijk04}). Fine-structure
levels can be populated through collisions, photo-excitation by IR photons,
and/or fluorescence (Bahcall \& Wolf~1968). Assuming that the former
mechanism is dominant (but see Berger et al.~2006; Prochaska et
al.~2006), we can estimate the \ion{H}{I} volume density, using the
calculations by Silva \& Viegas (2002).  Assuming an electron fraction
$n_e \sim 10^{-4} n_{\rm H I}$ (see Vreeswijk et al.~2004; Berger et
al.~2005) and a temperature of 1000~K, we find $n_{\rm H I} \sim
10^{5}$ cm$^{-3}$ for the fourth component, and $n_{\rm H I} \sim
10^{4}$ cm$^{-3}$ for the mean ratio of all components (see Fig.~8 of
Silva \& Viegas~2002). For the fourth component, we can actually
constrain the kinetic temperature and volume density, as the observed
ratio [\ion{O}{I*}/\ion{O}{I**}] $=0.78\pm0.17$ can only be reached
below 320 K (within the 1$\sigma$ errors), while
[\ion{Si}{II*}/\ion{Si}{II}] $= -0.34\pm0.11$ requires a temperature
above 240 K. This temperature range corresponds to a volume density of
$n_{\rm H I} = 1-3 \times 10^{7}$ cm$^{-3}$.

\begin{table}
\caption {Ionic column densities in individual components of the GRB system
at $z_{\rm abs}=4.048$.\label{metallines}}
\begin{center}
\begin{tabular}{llll}
\hline
\hline
Ion & Transition & $\log N\pm\sigma _{\log N}$ & $b\pm\sigma _b$\\
 & lines used &                             & (km s$^{-1}$)  \\
\hline
\multicolumn{4}{l}{$z_{\rm abs}=4.0441$}  \\
\ion{S}{ii}   & 1250, 1253, 1259 & 14.36$\pm$0.06 &  14$\pm$7 \\ 
\ion{Si}{ii*} & 1309             & $<13.10 ^{\rm a}$ &  14$\pm$7 \\ 
\hline
\multicolumn{4}{l}{$z_{\rm abs}=4.0480$}  \\
\ion{S}{ii}   & 1250, 1253, 1259 & 14.79$\pm$0.05 &  $\sim15$  \\ 
\ion{Si}{ii*} & 1309             & 13.54$\pm$0.08 &  $\sim15$  \\ 
\hline
\multicolumn{4}{l}{$z_{\rm abs}=4.0490$}  \\
\ion{S}{ii}   & 1250, 1253, 1259 & 14.60$\pm$0.06 &  $\sim15$  \\ 
\ion{Si}{ii*} & 1309             & 13.77$\pm$0.06 &  $\sim15$  \\ 
\hline
\multicolumn{4}{l}{$z_{\rm abs}=4.0509$}  \\
\ion{S}{ii}   & 1250, 1253, 1259 & 14.59$\pm$0.06 &  14$\pm$2  \\ 
\ion{Si}{ii*} & 1309             & 14.22$\pm$0.05 &  14$\pm$2  \\ 
\ion{O}{i*}   & 1304             & 15.03$\pm$0.16 &  14$\pm$2  \\ 
\ion{O}{i**}  & 1306             & 14.25$\pm$0.05 &  14$\pm$2  \\ 
\ion{Si}{ii}  & 1304             & 14.56$\pm$0.10 &  14$\pm$2  \\ 
\hline 
$^{\rm a}$ $3\sigma$ upper limit.
\end{tabular} 
\end{center} 
\end{table}     

Finally, our data show the first evidence for H$_2$ molecules in a GRB
absorber. Two consistent features are seen at the location of the W1-0
R(0), W1-0 R(1) and W1-0 Q(1) lines at $z_{\rm abs}=4.04793$ (bottom
panel in Fig.~\ref{LRMRspectrum}), with column densities of
log~$N$(H$_2$)~$\sim$~17.0 for the ${\rm J}=1$ rotational level, and
log~$N$(H$_2$)~$<$~16.7 for ${\rm J}=0$. The corresponding H$_2$
molecular fraction is $\log f\sim-3.5$ with
$f=2N($H$_2$)/(2$N$(H$_2$)+ $N(\ion{H}{i})$. This is the second
highest redshift at which these H$_2$ lines have been detected (Ledoux
et al.~\cite{ledoux06}).

\section{Discussion and conclusions} \label{dis.sec}

In Table~\ref{metallicitytable}, we have compiled metallicity
measurements for $z>2$ GRB absorption systems from the literature. As
seen, despite its high redshift, the GRB\,060206 system has one of the
highest metallicities measured for a GRB absorption system and one of
the highest metallicities measured from QSO absorption lines at
$z>4$. We also note that the metallicity of GRB\,060206 is about 15
times higher than that of GRB\,050730 which has a similar redshift
($z=3.968$). This shows that GRBs do not only occur in very low
metallicity environments, but also in environments covering a broad
range of metallicities at a given redshift. 

Fine-structure lines are ubiquitous in GRB absorbers (Vreeswijk et
al.~\cite{vreeswijk04}; Berger et al.~\cite{berger}; Chen et
al.~\cite{chen}). As discussed above, for GRB~060206 we derive high
densities and low temperatures, consistent with the possible detection
of H$_2$ in the spectrum, with a molecular fraction of $\log f
\sim-3.5$.  Molecules are also detected in QSO-DLAs (Ledoux et
al.~\cite{ledoux03}) in cold gas with T$\lesssim$ 300~K, but with
lower densities ($\lesssim400$ cm$^{-3}$, Srianand et
al.~\cite{srianand}). The difference in densities along random
(cross-section selected) and GRB sightlines like the one towards
GRB\,060206 must reflect the higher than average density of the star
forming region in which the progenitor star was located. Such regions
must have a cross-section which is less than a few percent of the
total cross-section for QSO-DLAs at redshifts $z=$ 2--4 in order to
explain that similar fine-structure lines have not yet been seen in
QSO-DLAs.

\begin{table}[t] 
\caption[]{Table of
published absorption metallicities for GRBs.
References: [1] Savaglio et al.~(2003); [2] Vreeswijk et al.~(2006); [3]
Vreeswijk et al.~(2004); [4] Watson et al.~(2006); [5] Berger et al.~(2005);
[6] Starling et al.~(2005); [7] Chen et al.~(2005); [8] Ledoux et al., 
2005; [9] Kawai et al.~(2005); [10] this work. \label{metallicitytable}} 
\begin{center} 
\begin{tabular}{lllr} 
\hline 
\hline 
GRB & Metallicity & Redshift & Ref \\ 
\hline 
000926 & [Zn/H]$ = -0.13\pm0.25$  & 2.038 & 1 \\ 
011211 & [Si/H]$ = -0.90\pm0.5$  & 2.142 & 2 \\ 
030323 & [S/H]$ = -1.26\pm0.2$  & 3.372 & 3 \\ 
050401 & [Zn/H]$ = -1.0\pm0.4$   & 2.899 & 4 \\
050505 & [S/H]$ \gtrsim -1.2$ &  4.275 & 5 \\ 
050730 & [S/H]$ = -2.0\pm0.2$  & 3.968 & 6,7 \\ 
050820 & [Si/H]$ = -0.6\pm0.1$  & 2.615 & 8 \\ 
050904 & [S/H]$ = -1.3\pm0.3$  & 6.295 & 9 \\ 
060206 & [S/H]$= -0.84\pm0.10$  & 4.048 & 10 \\ 
\hline 
\end{tabular} 
\end{center} 
\end{table}

In the collapsar models GRBs can only be formed by massive single stars
with a metallicity below $\sim$0.3 Z$_{\odot}$ (Hirschi et
al.~\cite{Hirschi}; Woosley \& Heger~\cite{WH06}).  Such a bias is not present
in all models, most noteably not in models involving a binary progenitor (Fryer
\& Heger \cite{fryer}). To gauge what a metallicity bias could mean for the
completeness of GRBs as cosmological probes it is important to know the present
mass-weighted mean metallicity and how it declines with redshift.  Zwaan et
al.~(\cite{zwaan}) find a present-day mean metallicity in the gas phase of
$\bar{Z}\approx0.44$ Z$_{\odot}$ and a slope between $-0.25$ and $-0.30$ dex
per unit redshift. This means that at $z\approx1$ and earlier the mean
metallicity of the gas is below the cut-off value above which single massive
stars in the collapsar models do not make GRBs. It is therefore likely that
GRBs at $z>2$ will be fairly unbiased tracers of star-formation, while they
become increasingly biased at $z<1$ if there is a low-metallicity bias. 

\begin{figure} 
\centering 
\epsfig{file=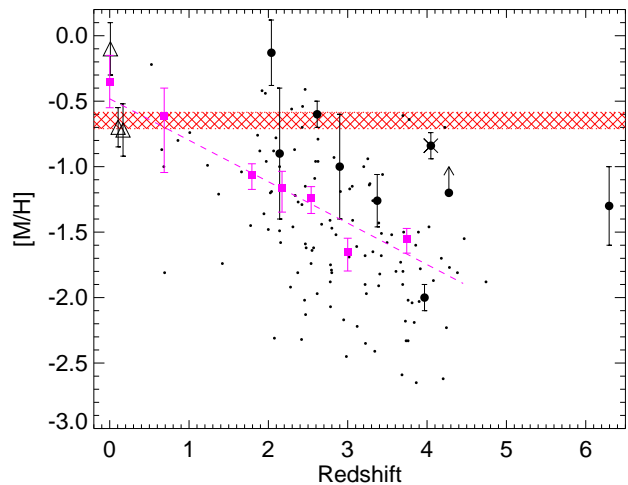,width=8.5cm,clip=}
\caption{Metallicity as a function of redshift for different classes
  of objects. The black circles are the measurements for GRBs from
  Table~\ref{metallicitytable} (GRB\,060206 is also marked with
  $\times$). The open triangles show the metallicity of three low-$z$
  GRB host galaxies (Sollerman et al. \cite{sollerman}).  The squares
  and the dashed line represent the column density weighted
  metallicity evolution derived by Zwaan et al.~(\cite{zwaan}, their
  Fig.~22). The small dots with no error-bars are measurements for 121
  DLAs from Prochaska et al.~(\cite{prochaska}). The hatched region
  indicates the metallicity above which GRBs cannot form in the
  collapsar models.
\label{Zplot}} 
\end{figure}

Metallicities can be measured from optical spectroscopy of the afterglow or
the host galaxy. Fig.~\ref{Zplot} shows that most metallicities
for GRB absorption systems fall below 0.3 Z$_{\odot}$, but one (GRB\,000926,
$Z=0.7$ Z$_{\odot}$, Savaglio et al.~\cite{savaglio}) has a metallicity well
above the threshold for a single massive star to produce GRBs in the collapsar
models. 
The metallicity of the host galaxy of GRB\,980425 by Sollerman et 
al.~(\cite{sollerman}) is also higher than this limit (0.8  Z$_{\odot}$). 
This indicates that collapsars resulting from single massive stars are
not the only progenitors to long GRBs or that massive stars with Z $>$ 0.3
Z$_{\odot}$ can also produce long GRBs. The GRB metallicities fall within the
range spanned by QSO-DLAs, but most are above the mean curve derived by Zwaan
et al.~(\cite{zwaan}). The mean offset is $0.49$ and the scatter 0.38 dex
(excluding GRB\,050904 which is at very high redshift where there are no 
QSO-DLA data).
This offset may reflect that DLAs are cross-section selected,
whereas GRBs represent sightlines to more central regions in their hosts (Bloom
et al.~\cite{bloom}). A metallicity gradient of $-0.09$ dex kpc$^{-1}$ as
assumed in the study by Zwaan et al.~(\cite{zwaan}) (and observed in the
Galaxy) seems sufficient to explain the offset. Note that the current sample
of QSO-DLAs is not likely to be strongly biased against high metallicity 
sightlines (Akerman et al.~\cite{akerman}).

Furthermore, in the derivation of the cosmic
star-formation density, and hence the production of metals, the shape of the
luminosity function is important.  Roughly, two thirds of the UV light from
LBGs is emitted by galaxies that are fainter than the actual flux limit of the
ground based LBG survey of Steidel et al. (2003). Using only LBGs to derive the
total star-formation density means that an extrapolation to the poorly
determined faint end of the luminosity function is unavoidable. GRBs allow us
to probe this faint end since the selection is not limited by the brightness of
the host. We refer to Jakobsson et al.~(\cite{palli05}) for a quantitative
analysis. 
Finally, GRBs allow the measurement of metallicities at very
high redshifts ($z>6$, \cite{Kawai}), which is inaccessible to QSO-DLAs.

In conclusion, we measure a metallicity of [S/H]$=-0.84\pm0.10$ for GRB\,060206
at $z=4.048$. We also find the first evidence for molecular lines
from a GRB absorber. We have argued that GRBs can be used to measure the
metallicities and luminosities of typical star forming galaxies at $z>2$,
making GRBs promising complementary probes of chemical evolution at high
redshift.

\begin{acknowledgements} 
The authors acknowledge the indispensable assistance given by both
observers and staff at WHT and NOT. We also thank C. P{\'e}roux,
S. Ellison, and J. Andersen for helpful discussions. The Dark
Cosmology Centre is funded by the Danish National Research
Foundation. KW and PC thank NWO for support under grant
639.043.302. AL, NRT and ER thank PPARC for support.  The research of
JG is supported by the Spanish Ministry of Science and Education
through programmes ESP2002-04124-C03-01 and AYA2004-01515. JMCC
acknowledges partial support from IDA and the NBI's International
Ph.D. School of Excellence. We also acknowledge benefits from
collaboration within the EU FP5 Research Training Network ``Gamma-Ray
Bursts: An Enigma and a Tool" (HPRN-CT-2002-00294).
\end{acknowledgements}


\begin{thebibliography}{}

\bibitem[2005]{akerman} Akerman, C. J., Ellison, S. L., Pettini, M., \&
Steidel, C. C. 2005, A\&A, 440, 499

\bibitem[1968]{bahcall} Bahcall, J. N. \& Wolf, R. A. 1968, ApJ, 152, 701

\bibitem[2006]{band} Band, D. L. 2006, ApJ, in press, astro-ph/0602267

\bibitem[2005]{berger} Berger, E., Penprase, B. E., Cenko, S. B., et al. 2005,
ApJ, submitted (astro-ph/0511498)

\bibitem[2002]{bloom} Bloom, J. S., Kulkarni, S. K., \& Djorgovski, S. G. 
2002, AJ, 123, 1111

\bibitem[2006]{boyd} Boyd, P., Morris, D., Marshall, F., Gehrels, N.  2006, GCN
Circ. 4722 

\bibitem[2005]{chen}Chen, H.-W., Prochaska, J. X., Bloom, J. S., \& Thompson,
I. B. 2005, ApJ, 634, L25

\bibitem[2005]{ferrara} Ferrara, A., Scannapieco, E. \& Bergeron, J. 2005, ApJ,
634, L37

\bibitem[2005]{fryer} Fryer, C. L., \& Heger, A. 2005, ApJ, 623, 302

\bibitem[2004]{gehrels} Gehrels, N., Chincarini, G., Giommi, P., et al.  2004,
ApJ, 611, 1005

\bibitem[2003]{heger} Heger, A., Fryer, C. L., Woosley, S. E., Langer, N., \&
Hartmann, D. H. 2003, ApJ, 591, 288

\bibitem[2005]{Hirschi} Hirschi, R., Meynet, G., \& Maeder, A. 2005, A\&A, 443,
581

\bibitem[Hjorth et al.~2003]{jensNature} Hjorth, J., Sollerman, J., M\o ller,
P., et al.  2003, Nature, 423, 847

\bibitem[2005]{palli05} Jakobsson, P., Bj\"ornsson, G., Fynbo, J. P. U., et al.
2005, MNRAS, 362, 245

\bibitem[Jakobsson et al.~2006]{palli06} Jakobsson, P., Levan, A., Fynbo, J. P.
U., et al. 2006, A\&A, 447, 897

\bibitem[Kawai et al.~2006]{Kawai} Kawai, N., Kosugi G., Aoki, K., et al.,
2005, Nature 440, 184

\bibitem[2003]{ledoux03} Ledoux, C., Petitjean, P., \& Srianand, R. 2003,
MNRAS, 346, 209

\bibitem[2005]{ledoux} Ledoux, C., Vreeswijk, P., Ellison, S., et al. 
2005, GCN Circ. 3860

\bibitem[2006]{ledoux06} Ledoux, C., Petitjean, P., \& Srianand, R. 2006,
ApJL, in press, astro-ph/0602212

\bibitem[2006a]{morris} Morris, D. C., Barbier, L., Barthelmy, S. 2006a, GCN
Circ. 4682 

\bibitem[2006b]{morrisb} Morris, D. C., Burrows, D., Gehrels, N., Boyd, P., \&
Angelini, L. 2006b, GCN Circ. 4689 

\bibitem[2006]{Palmer}Palmer, D., Barbier, L., Barthelmy, S., et al.  2006, GCN
Circ. 4697

\bibitem[1995]{peifall} Pei, Y. C., \& Fall, M. S. 1995, ApJ 454, 69

\bibitem[2003]{prochaska}Prochaska, J. X., Gawiser, E., Wolfe, A. M., Castro,
S., \& Djorgovski, S. G. 2003, ApJ, 595, L9

\bibitem[2006]{prochaska06}Prochaska, J. X., Chen, H.-W. \& Bloom, J.~S. 2006, 
ApJ, submitted, astro-ph/0601057

\bibitem[2002]{sv02} Silva, A. I., \& Viegas, S. M. 2002, MNRAS, 329, 135

\bibitem[2005]{sollerman} Sollerman, J., {\"O}stlin, G., Fynbo, J. P. U., et al.
2005, NewA, 11, 103

\bibitem[2005]{srianand} Srianand, R., Petitjean, P., Ledoux, C., 
Ferland, G., \& Shaw, G. 2005, MNRAS, 362, 549

\bibitem[Stanek et al. 2003]{stanek03}
Stanek, K. Z., Matheson, T., Garnavich, P. M., et al. 2003, \apj, 591, L17

\bibitem[2005]{rhaana} Starling, R. L. C., Vreeswijk, P. M., Ellison, S. L., et
al. 2005, A\&A, 442, L21

\bibitem[2003]{savaglio} Savaglio, S., Fall, S. M., \& Fiore, F. 2003, ApJ,
585, 638

\bibitem[2004]{vreeswijk04} Vreeswijk, P. M., Ellison, S. L., Ledoux, C., et
al. 2004, A\&A, 419, 927

\bibitem[2006]{vreeswijk06} Vreeswijk, P. M., Smette, A., Fruchter, A. S., et
al. 2006, A\&A, 447, 145

\bibitem[Wijers et al.~1998]{wijers98} Wijers, R. A. M. J., Bloom, J. S., 
Bagla, J. S., \& Natarajan, P.  1998, MNRAS, 297, L13

\bibitem[2006]{WH06} Woosley, S. E., \& Heger, A. 2006, ApJ, 637, 914

\bibitem[2005]{zwaan} Zwaan, M. A., van der Hulst, J. M., Briggs, F. H.,
Verheijen, M. A. W., \& Ryan-Weber, E. V. 2005, MNRAS, 364, 1467

\end{thebibliography}
\end{document}